\documentclass[10pt,prx,twocolumn,amsmath,amssymb,floatfix,notitlepage,superscriptaddress]{revtex4-2}
\usepackage[T1]{fontenc}
\usepackage[utf8]{inputenc}
\usepackage{lmodern}
\usepackage{microtype,bm,bbm}
\usepackage{graphicx,booktabs}
\usepackage{times}
\usepackage[usenames,dvipsnames]{xcolor}
\usepackage[english]{babel}
\usepackage{upgreek}
\usepackage{braket}
\usepackage{subfigure}
\usepackage{dsfont}
\usepackage{transparent}
\usepackage{layouts}
\usepackage{tikz}
\usepackage{adjustbox}

\usepackage{printlen}

\usepackage{hyperref}
\hypersetup{
  colorlinks,
  allcolors=NavyBlue,
  linktoc=all
}

\usepackage[capitalize]{cleveref}
\usepackage{pifont}
\crefformat{section}{Sec.~#2#1#3}

\graphicspath{ {figures/}}
\graphicspath{ {inkscape_figures/}}

\begin{document}

\title{Measurement-free quantum error correction optimized for biased noise}

\author{Katharina Brechtelsbauer}
\affiliation{Institute for Theoretical Physics III and Center for Integrated Quantum Science and Technology, University of Stuttgart, Pfaffenwaldring 57, 70569 Stuttgart, Germany }

\author{Friederike Butt}
\affiliation{Institute for Quantum Information, RWTH Aachen University, Aachen, Germany}
\affiliation{Institute for Theoretical Nanoelectronics (PGI-2), Forschungszentrum J\"{u}lich, J\"{u}lich, Germany}

\author{David F. Locher}
\affiliation{Institute for Quantum Information, RWTH Aachen University, Aachen, Germany}
\affiliation{Institute for Theoretical Nanoelectronics (PGI-2), Forschungszentrum J\"{u}lich, J\"{u}lich, Germany}

\author{Santiago Higuera Quintero}
\affiliation{Institute for Theoretical Physics III and Center for Integrated Quantum Science and Technology, University of Stuttgart, Pfaffenwaldring 57, 70569 Stuttgart, Germany }

\author{Sebastian Weber}
\affiliation{Institute for Theoretical Physics III and Center for Integrated Quantum Science and Technology, University of Stuttgart, Pfaffenwaldring 57, 70569 Stuttgart, Germany }

\author{Markus M\"{u}ller}
\affiliation{Institute for Quantum Information, RWTH Aachen University, Aachen, Germany}
\affiliation{Institute for Theoretical Nanoelectronics (PGI-2), Forschungszentrum J\"{u}lich, J\"{u}lich, Germany}

\author{Hans Peter Büchler}
\affiliation{Institute for Theoretical Physics III and Center for Integrated Quantum Science and Technology, University of Stuttgart, Pfaffenwaldring 57, 70569 Stuttgart, Germany }

\date{\today}

\begin{abstract}

In this paper, we derive  optimized  measurement-free protocols for quantum error correction and the implementation of a universal gate set optimized for an error model that is noise biased . The noise bias is adapted for neutral atom platforms, where 
two- and multi-qubit gates are realized with Rydberg interactions and are thus expected to be the dominating source of noise. Careful design of the gates allows to further reduce the noise model to Pauli-Z errors. 
In addition, the presented circuits are robust to arbitrary single-qubit gate errors, and  we demonstrate that the break-even point can be significantly improved compared to fully fault-tolerant measurement-free schemes.  The obtained logical qubits with their suppressed error rates on logical gate operations 
can then be used as building blocks in a first step of error correction in order to push the effective error rates below the threshold of a fully fault-tolerant and scalable quantum error correction scheme.  

\end{abstract}
\maketitle

\section{Introduction}

Achieving fault tolerance is one of the key challenges in the field of quantum computing. Quantum error correction (QEC) is a promising tool to protect quantum information against noise \cite{Campbell2016}. Recently, great progress has been made in the implementation of QEC codes on various quantum computing platforms \cite{Ryan-Anderson2024,Putterman2024,Pogorelov2024,Bluvstein2024,Rodriguez2024,Acharya2024}. 
Fault-tolerant quantum computing requires the ability to encode logical qubits, to correct errors, and to perform logical gates, while suppressing the uncontrolled spread of errors during these operations.  
However, a QEC code with a transversal universal logical gate set cannot exist \cite{Eastin2009}. Promising candidates for the completion of universal logical gate sets are for example magic state injection \cite{Bravyi2005,Goto2016} and code switching \cite{Anderson2014,Bombin2016}. However, most of these implementations as well as many QEC protocols require mid-circuit measurements and feed-forward operations. On many platforms, measurements are slow and thus induce high error rates on idling qubits. There are several proposals to speed up measurements, like enhancing the photon signal using optical cavities \cite{Hu2024,Grinkemeyer2024} or by means of several auxiliary qubits \cite{Saffman2005,Petrosyan2024}. An alternative is measurement-free QEC \cite{Aharonov2008,PazSilva2010,Boykin2010,Crow2016,Perlin2023,Goto2023,Ouyang2024,Gottesman2024}. Here, instead of measuring auxiliary qubits and applying feedback depending on the measurement outcome, the coherent feedback is directly implemented with quantum gates controlled by the auxiliary qubits. Entropy is then removed from the system by resetting the auxiliary qubits afterwards. Fault-tolerant circuits for error correction without mid-circuit measurements have been proposed in \cite{Heussen2024,Veroni2024}, and recently, measurement-free, fault-tolerant  and scalable universal logical gate sets have been constructed \cite{Butt2024,Veroni2024_2}. Although measurement-free QEC does not require mid-circuit measurements, it comes with the cost of additional quantum gates in the feedback and thus higher gate fidelities are required. To reduce the requirements and optimize the threshold, it is beneficial to tailor the QEC protocols towards the characteristics of the respective platform \cite{Tuckett2018,Tuckett2019,Cong2022}.

One promising platform for quantum computing are neutral atoms that are trapped in optical tweezers \cite{Henriet2020,Graham2022,Ma2022,Unnikrishnan2024,Pucher2024,Cao2024,Pichard2024,Reichardt2024,Manetsch2024,Bluvstein2024,Rodriguez2024}. Tweezer arrays allow one to arrange atoms in flexible geometries and offer the possibility to re-shuffle atoms during the computation, which implies high connectivities \cite{Bluvstein2022}. Furthermore, the arrays can be scaled up to large numbers of qubits, and recently arrays with thousands of qubits have been demonstrated \cite{Manetsch2024}. The qubits are encoded in long-lived electronic states, for example different hyper-fine ground states or clock states. High-fidelity single-qubit gates can be implemented using laser or microwave pulses. Entangling gates make use of the Rydberg blockade mechanism, where the excitation of one atom to a Rydberg state blocks the excitation of neighboring atoms \cite{Jacksch2000}. Furthermore, the blockade mechanism natively supports the implementation of multi-qubit gates. In the past years, great progress has been made in achieving high-fidelity single- and two-qubit gates \cite{Levine2022,Ma2022,Unnikrishnan2024,Pucher2024,Cao2024,Levine2019,Evered2023}. Typical sources of noise are laser noise, photon recoil and decay from the Rydberg state. Careful gate design allows to simplify the physical noise model, for example by converting all errors to erasure errors \cite{Wu2022,Scholl2023} or to phase errors \cite{Cong2022,Sahay2023}. 

In this work, we construct reduced QEC circuits and a universal logical gate set  for a measurement-free setup, where Z-errors on two- and multi-qubit gates dominate the noise model.   
This biased noise model allows us to decrease the qubit and gate overhead and thus push the break-even point (where logical errors become less likely than physical errors) into a regime that is feasible with state of the art experiments. 
We encode the logical qubits in $[[7,1,3]]$ Steane codes, which allows for the transversal realization of Clifford gates. To implement the T-gate we invent protocols for measurement-free magic state injection that are fault-tolerant under the biased noise model.
Remarkably, we find that the protocols are even robust to arbitrary errors on single-qubit gates. Furthermore, we show that the discussed protocols can still be used to suppress the logical error rate even if the noise is not perfectly biased and bit-flip errors on two- and multi-qubit gates occur with a small probability. 

The 7-qubit Steane code allows the correction of up to one physical error. The general idea is that such a first round of error correction optimized for a platform-specific noise model allows one to reduce the error rate of the logical operations below the threshold of a scalable and fully fault-tolerant error correction protocol.
The obtained logical qubits with their suppressed error rates can then be used as building blocks for more general QEC codes, where physical error rates lie above the threshold, to perform universal quantum computation.

The rest of this work is structured as follows. In \cref{sec:setup} we review the 7-qubit Steane code and introduce the biased noise model. In \cref{sec:logicalqubits} we present the reduced QEC circuits and the T-gate and discuss the logical noise model of all relevant quantum operations. In \cref{sec:imperfectnoisebias} we consider more general noise models and discuss the effects of imperfect noise bias. In \cref{sec:implementation} we discuss possible implementations on Rydberg platforms. Finally, we conclude in \cref{sec:conclusions}.

\section{Setup}
\label{sec:setup}

\begin{figure}[tb]
  \includegraphics[width=0.8 \linewidth]{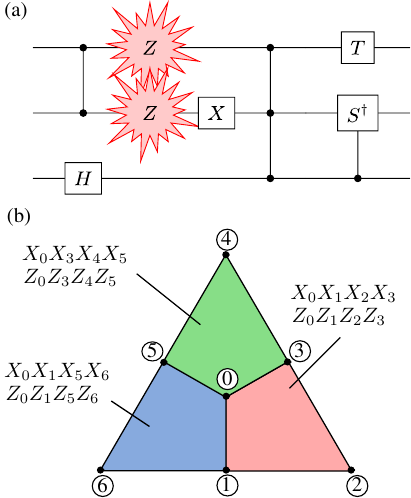}
  \caption{(a) Exemplary circuit. Our set of basis gates includes controlled-phase gates (in particular the CZ-gate and CS$^\dagger$-gate), H-, X- and T-gates and multi-controlled Z-gates (for example the CCZ-gate). Initially, we assume the noise model defined in \cref{eq:biasednoise}, where only Z-type errors appear after two- and multi-qubit gates. (b) The $[[7,1,3]]$-qubit Steane code. The code has distance $d=3$ and is designed to correct bit-flip errors and phase errors on up to one qubit. The code is defined by the stabilizer group generated by the operators shown. The logical $X_L$ and $Z_L$ operators act on one side of the triangle.}
  \label{fig:setup}
\end{figure}

Our goal is to provide logical qubits with reduced error rates that can be used as building blocks for higher level error correction protocols or to directly perform quantum algorithms. This requires the ability to encode information into the logical subspace, QEC protocols and a universal logical gate set. While all these building blocks are usually built on  mid-circuit measurements, we aim to provide measurement-free protocols. 

As logical qubits we use the 7-qubit Steane code \cite{Steane1996}.
The $[[7,1,3]]$ Steane code encodes $k=1$ logical qubit into $n=7$ physical qubits. The code has distance $d=3$ and is designed to independently correct bit-flip errors and phase errors on up to one qubit.
The code is defined by the stabilizer group generated by the operators (see \cref{fig:setup})
\begin{align}
  X_0X_1X_2X_3, \quad Z_0Z_1Z_2Z_3, \notag \\
  X_0X_3X_4X_5, \quad Z_0Z_3Z_4Z_5, \\
  X_0X_1X_5X_6, \quad Z_0Z_1Z_5Z_6.  \notag
\end{align}
 The X-stabilizers enable the correction of Z-errors and the Z-stabilizers enable the correction of X-errors. The logical operators $X_L=X_1X_2X_6$ and $Z_L=Z_1Z_2Z_6$ act on one side of the triangle  (see \cref{fig:setup}). A convenient property of the Steane code is that many logical gates can be performed transversally, in particular the logical H-gate is realized by applying a physical H-gate to every single data qubit, the logical S-gate is realized by applying a physical S$^\dagger$-gate to every data qubit and the logical CZ-gate is obtained by applying physical CZ-gates to every pair of data qubits in the two code blocks. The only gate that is missing for universality is the T-gate, which can be implemented by means of magic state injection or code switching.

A compact circuit for fault-tolerantly encoding $\ket{0_L}$ without mid-circuit measurements is given in \cite{Heussen2024} and fault-tolerant QEC-cycles without mid-circuit measurements are presented in \cite{Heussen2024,Veroni2024}. As already mentioned, for the universal gate set on the Steane code one only needs to implement a T-gate, as all other gates needed for universality can be performed transversally. In \cite{Butt2024}, a measurement-free fault-tolerant T-gate for the Steane code is presented based on code switching. Here, we consider a platform-specific noise model, to reduce the requirements for the error correction cycles and the T-gate.

Inspired by Rydberg quantum processors \cite{Cong2022,Sahay2023}, we consider a platform that natively supports single-qubit rotations (in particular X-,H-,S-, and T-gate) as well as multi-controlled Z-gates (CZ, CCZ, CCCZ) and controlled-phase gates (CP). Initially, we assume that the only gates that can fail are the two- and multi-qubit gates with noise channels
\begin{align}
\mathcal{E}_l(\rho)=(1-p) \rho + \frac{p}{2^l-1} \sum_{P \in \{I,Z\}^{\otimes l} \backslash I^{\otimes l}} P\rho P,
\label{eq:biasednoise}
\end{align}
where $l$ is the number of qubits involved in the gate and $p$ the error rate. The noise model is illustrated exemplarily in \cref{fig:setup} and a detailed discussion of the validity of this noise model is given in \cref{app:detailsnoisemodel}. In \cref{sec:imperfectnoisebias} we extend our results to more general noise models including faulty single-qubit gates, faulty initialization and bit-flip errors after two- and multi-qubit gates. 

The biased noise model suggests that it is sufficient to use a QEC-code for correcting Z-errors only. However, with the above gate set Z-errors can be converted into X-errors that are then not correctable anymore. For example, a CX-gate with the assumed gate set would be compiled into two H-gates and a CZ-gate. If the CZ-gate fails and a Z-error occurs, the H-gate rotates the Z-error into an X-error. Thus, we consider the 7-qubit Steane code that can correct X-errors as well. The benefit of the biased noise model enters in form of reduced requirements for fault-tolerant error correction cycles and the logical T-gate.

\section{Logical qubits with biased noise}
\label{sec:logicalqubits}

\begin{figure*}[tb]
  \includegraphics[width= \textwidth]{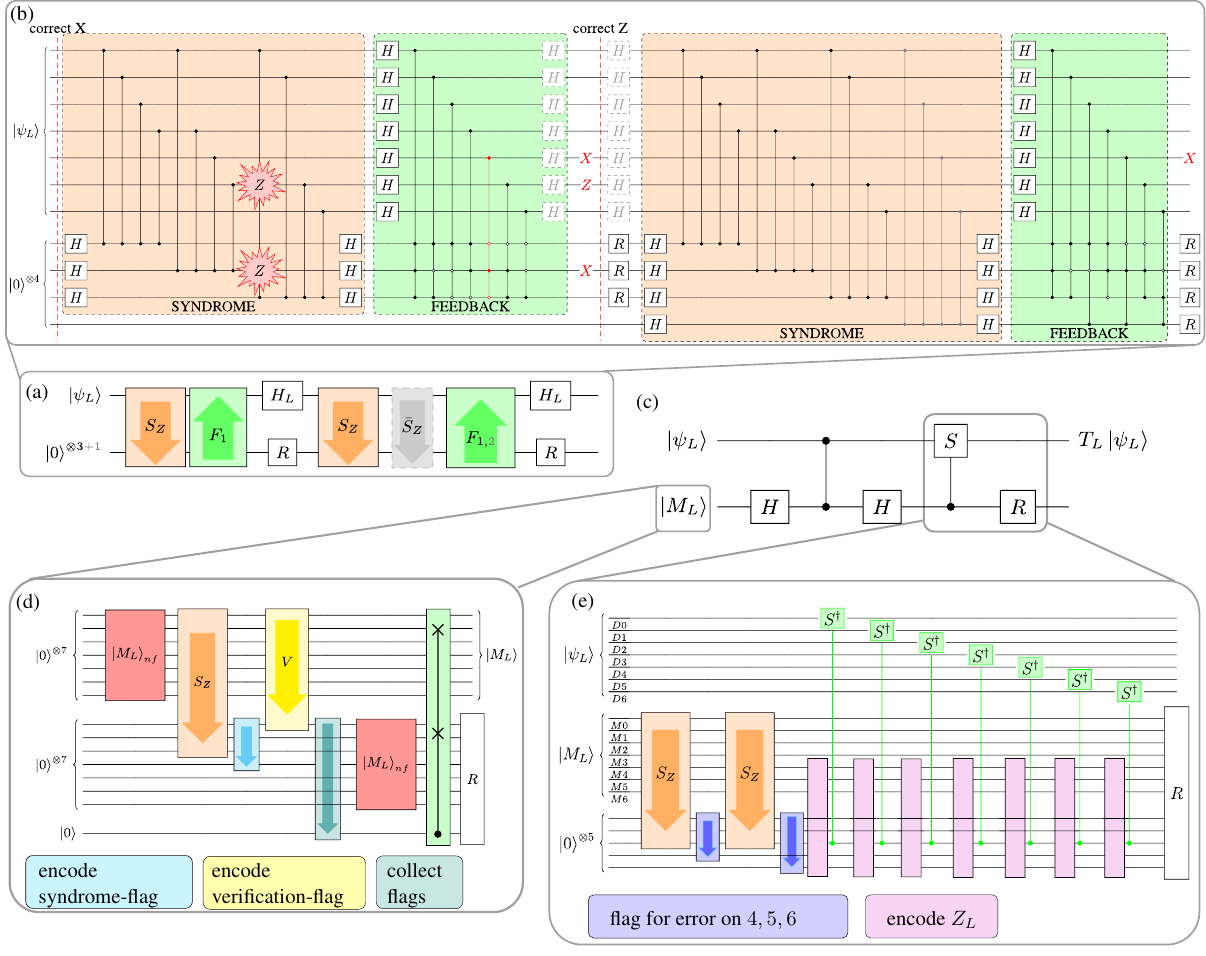}
  \caption{Circuits for error correction and T-gate. (a) Schematic circuit for the error correction scheme. For the second round of stabilizer extraction an additional stabilizer $X_0X_2X_4X_6$ (gray gates) has to be mapped out to avoid combinations of X- and Z-errors. However, this is only required before performing logical T- and S-gates. Otherwise, the extraction of an overcomplete set is not required and the feedback for correcting Z-errors is similar to the feedback for correcting $X$ errors (first green box). We refer to this as reduced error correction cycle. (b) Detailed error correction circuit. (c) Magic state injection. (d) Logical encoding of the magic state. (e) Logical CS-gate with subsequent reset on the control. The detailed circuits for the encoding of the magic state and the logical CS-gate with subsequent reset are shown in \cref{app:circuits}. The operation $R$ describes qubit reset. }
  \label{fig:main_circuit}
\end{figure*}

In this section we first present the measurement-free circuits for encoding of $\ket{0_L}$, error correction and the T-gate and in particular show that the requirements for the error correction cycle and the implementation of the T-gate can be reduced for the biased noise model.
Afterwards, we  discuss the logical noise model of all building blocks and compare the measurement-free QEC-cycle and  T-gate with measurement-based schemes in the presence of idling noise.
  
\subsection{Encoding of $\ket{0_L}$}

We use the circuit presented in reference \cite{Heussen2024} to encode the state $\ket{0_L}$ in a measurement-free fashion. The encoding circuit is shown in the appendix in \cref{fig:full_encoding_logical_zero}. The logical zero state is initially encoded non-fault-tolerantly. Note, while in general weight-2 errors cannot be corrected in the $[[7,1,3]]$ Steane code, this is possible when encoding the $\ket{0_L}$-state, since here the  desired logical state is known. To detect potential weight-2 errors, the operator $Z_L$ and the stabilizer $Z_0 Z_2 Z_4 Z_6$ are subsequently mapped out into auxiliary qubits. If a dangerous weight-2 error occurred, both auxiliary qubits are flipped. In this case, the qubit in the middle of the Steane-triangle has to be flipped, such that at most a weight-1 error is left. To correct the error without measurement, one applies a CCX-gate, controlled by the two auxiliary qubits.  The circuit is fault-tolerant with respect to a general depolarizing noise model. However, also for the biased noise model weight-2 errors during the non-fault-tolerant encoding can appear, making the $Z_L$- and stabilizer extractions indispensable.

\subsection{Error correction cycle with biased noise}

The reduced error correction cycle for the Steane code with biased noise is illustrated in \cref{fig:main_circuit}. We verify numerically that the QEC-protocol is fault-tolerant with respect to the biased noise model in \cref{eq:biasednoise} by simulating all possible error patterns, where one gate fails (see also \cref{app:numerics}). In the Steane code X- and Z-errors can be corrected independently. Thus, we first extract the Z-stabilizers to detect X-errors and apply a coherent feedback operation to correct them. To correct Z-errors, we then apply a logical H-gate (a physical H-gate to all data qubits) to transform Z-errors into X-errors and apply again the circuit for correcting X-errors. Finally, another logical H-gate is applied to map the state back to the initial logical state. The extraction of an overcomplete set of stabilizers in the second half of the protocol ensures that a single error on the syndrome qubits cannot lead to wrong feedback, such that a faulty error correction cycle cannot introduce combinations of X- and Z-errors. In the first half of the error correction protocol such errors can appear if a CZ-gate fails on both qubits during the stabilizer extraction. Such a fault leads immediately to a Z-error on one data qubit and in addition to an X-error after the feedback (propagated by the gate marked in red in panel (b) in \cref{fig:main_circuit}). However, in the second half of the protocol the Z-error is corrected, such that the remaining error is a weight-1 X-error.

It is important to point out that the Steane code allows one to correct X- and Z-errors independently and combinations of X- and Z-errors are thus in principle correctable. However, a logical S- or T-gate transforms such an error combination into a different weight-2 error that is not correctable. Therefore, we have to extract the overcomplete set of stabilizers before performing a logical phase-gate.
This implies that the circuit can be shortened by removing the additional stabilizer extraction if no S- or T-gate follows before the next error 
correction round. In this shortened circuit, the feedback for correcting Z-errors is similar to the feedback for correcting X-errors. In the following, we refer to this as \textit{reduced error correction cycle}. 

Typically, when considering more general noise models like the depolarizing noise model, error correction protocols come along with more overhead. Here, we give a brief overview why our error correction protocol is applicable for biased noise:  One typical kind of error preventing fault-tolerance are hook errors, where one syndrome qubit is flipped during the syndrome extraction and the error then spreads to several data qubits via the remaining entangling gates in the syndrome extraction. However, since the biased noise model gives only rise to Z-errors after two-qubit gates, hook errors cannot occur. Similarly, failures during the feedback on the syndrome qubits do not have any effect, since Z-type errors commute with CCCZ gates. Another typical error-type are errors during the syndrome extraction on data qubits that lead to erroneous corrections. If for example an X-error happens on data qubit 0 during the extraction of the first Z-stabilizer this triggers the remaining stabilizers, but the first stabilizer is not affected, which leads to an erroneous correction of data qubit 5 and thus a weight-2 error. However, for the biased noise model the only errors that can happen during the extraction of the stabilizers are Z-errors, which do not trigger Z-stabilizers. 

\subsection{T-gate with magic state injection}

One way to implement a logical T-gate on the Steane code is magic state injection. Here, the idea is to prepare a logical auxiliary qubit in a magic state 
\begin{align}
\ket{M_L}=(\ket{0_L}+\exp(i \pi/4) \ket{1_L})/\sqrt{2},
\end{align}
that has imprinted the desired $\pi/4$-phase. The phase is then injected onto the logical qubit by entangling it with the magic state and after measuring the auxiliary logical qubit applying a logical S-gate depending on the outcome. This approach is especially suitable for the Steane code with the transversal S-gate. To perform magic state injection in a measurement-free manner, one can in principle replace the measurement with subsequent S-gate by a logical CS-gate \cite{Boykin2010}. However, the logical CS-gate is not transversal. Thus, to perform a logical T-gate, we have to find a way to encode the magic state in a measurement-free manner and to implement a measurement-free CS-gate. Since the magic state is thrown away after the injection, we can relax the requirements on the measurement-free CS-gate. The schematic circuit for the magic state preparation and the logical CS-gate with subsequent reset on the control qubit is shown in figure  \cref{fig:main_circuit}, the detailed circuits are shown in \cref{app:circuits}. 

To fault-tolerantly encode the magic state, we initially encode it non-fault-tolerantly (red box in \cref{fig:main_circuit} (d)). This is then followed by the extraction of all three Z-stabilizers (orange box) and a verification step (yellow box). The verification ensures that the prepared state is the magic state $\ket{M_L}$ and not the orthogonal state $Z_L \ket{M_L}$. If the state $Z_L \ket{M_L}$ was prepared, the verification-flag would be raised. The extraction of Z-stabilizers is needed, since pairs of weight-1 X- and Z-errors that can appear during the non-fault-tolerant encoding, are transformed into non-correctable errors by the verification. 
In a measurement-based scheme one would now measure the syndromes and the verification-flag, and then prepare a new magic state if the verification flag was raised or an X-error was detected. To make this measurement-free, we always encode a second magic state in a non-fault-tolerant manner, analogously to the first magic state preparation. Furthermore, the information if any X-error occured is encoded into a flag (syndrome-flag). Finally, we collect the information of the syndrome-flag and the verification-flag into one new flag that is used to control a logical SWAP-gate on the two magic states. If the syndrome extraction or the verification detects an error, the two magic states are swapped and the second magic state is used for the T-gate. The logical SWAP-gate with single-qubit control can be implemented with logical CCZ- and H-gates. To reduce the number of required qubits, we first encode the syndrome- and verification-flags and prepare the final control-flag and afterwards encode the second magic state as illustrated in \cref{fig:main_circuit}.

To perform the CS-gate with subsequent reset, we encode the logical Z-operator $Z_L=Z_4Z_5Z_6$ of the logical magic state qubit into auxiliary qubits (pink boxes in \cref{fig:main_circuit} (e)), which are then used to control physical CS-gates \cite{Boykin2010}. To ensure fault-tolerance, every CS-gate is controlled by a freshly encoded auxiliary qubit. In addition, there are two rounds of Z-stabilizer measurements (orange boxes) before extracting the logical Z-operator, to detect errors on the qubits 4, 5 and 6. If such an error is detected, the information on the auxiliary qubit is flipped (included in extraction of logical Z in \cref{fig:main_circuit}). Resetting one of the auxiliary qubits collapses the magic state into the state $\ket{0_L}$ or $\ket{1_L}$. However, since the magic state injection anyway requires only a classical control, this does not affect the successful implementation of the T-gate.

Again, fault-tolerance of the T-gate with magic state injection with respect to the biased noise model in \cref{eq:biasednoise} is verified numerically.

A promising alternative to achieve a logical universal gate set is code switching, where one switches between codes with complementary gates sets. As a comparison we simplify the measurement-free code switching protocols presented in \cite{Butt2024} to the biased noise model in \cref{eq:biasednoise}. However, we find that with the noise bias a T-gate based on reduced code switching has a substantially lower break-even point than the T-gate based on our scheme for magic state injection. Thus, we focus on the magic state injection. Nevertheless, the reduced code switching can be useful for example for the implementation of logical CCX-gates, which require only one round of code switching on every involved logical qubit, but at least four T-gates and thus four rounds of magic state injection.

\subsection{Logical noise model}

\begin{figure}[tb]
  \includegraphics[width= \linewidth]{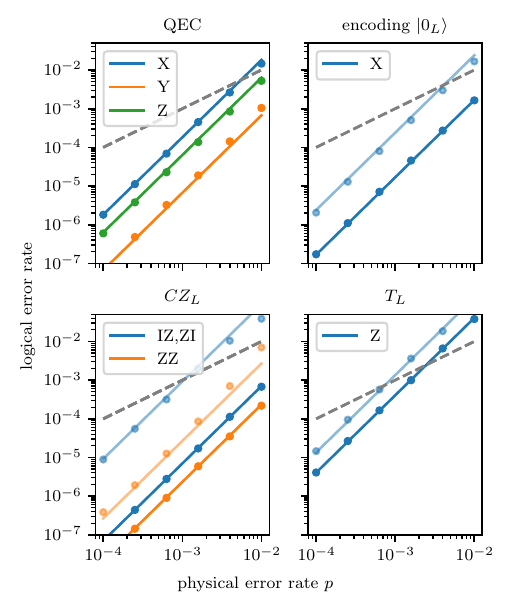}
  \caption{Characterization of logical noise model for the error correction cycle, encoding of $\ket{0_L}$, a logical CZ-gate and a logical T-gate. Here and in the following, the dashed line represents the physical error rate $p$, to indicate the break-even point. For the encoding and the logical gates the dark lines refer to the bare gate and the light lines to the gate surrounded by error correction cycles (extended rectangles). The points show the numerically simulated data and lines are the analytical estimates for the logical error rates (see \cref{app:lognoisemodel}). The bare logical CZ- and T-gate can only give rise to Z-errors. In principle, for the logical CZ-gate and the T-gate surrounded by QEC one also finds X- and Y-type errors. X- and Y-type errors for the extended CZ- and T-gate and the noise model for extended H- and S-gate are shown in \cref{app:lognoisemodel}.}
  \label{fig:logicalnoise}
\end{figure}

Next, we analyze the effective noise of the logical building blocks (gadgets). To find the effective errors on the logical qubit, we randomly place errors in a gadget using Monte-Carlo Sampling (see \cref{app:numerics}). We then perform the faulty gadget on the test state $(\ket{0_L}+e^{ i \pi/8} \ket{1_L})/\sqrt{2}$ for single-qubit gadgets and $(\ket{0_L0_L}+e^{ i \pi/8} \ket{1_L0_L}+e^{ i \pi/8} \ket{0_L1_L}-e^{ i \pi/4} \ket{1_L1_L})/2$ for two-qubit gadgets. Afterwards, we compare the final state with the state one obtains after performing the ideal gadget and deduce the Kraus operators. Note that in general to find the Kraus operators one has to perform state tomography. However, since we only place Pauli errors, the possible final errors in the considered circuits are  X-,Y-,Z- and S-type errors, which can all be clearly distinguished using the above state. The rates of these errors for the QEC cycle, the CZ-gate, the encoding of $\ket{0_L}$ and the T-gate is shown in dark colors in \cref{fig:logicalnoise}. In the following, we first discuss the logical noise model of the error correction circuit and the summarize the main findings for the encoding, the logical CZ-gate and the logical T-gate.

We find that the error correction cycle gives rise to X-, Y- and Z-type errors. During the correction of X-errors the extraction of the Z-stabilizers can lead to Z-errors while the feedback can result in X-errors. During the correction of Z-errors the stabilizer extraction gives rise to X-errors and the feedback can lead to Z-errors. Since we extract one more X-stabilizer than Z-stabilizers, the probability for X-errors during syndrome extraction is higher than the probability for Z-errors. Furthermore, if an overcomplete set of stabilizers is used for the feedback an error on a syndrome cannot lead to wrong corrections on data qubits. Thus, the probability to have X-errors is higher than the probability for Z-errors. Logical Y-errors can only appear, if there are at least two X-errors and two Z-errors on the data qubits. In leading order ($p^2$) this is only the case if one CZ-gate fails during the extraction of the Z-stabilizers on both qubits and one CZ-gate during the extraction of the X-stabilizers on both qubits. As discussed before, overcomplete stabilizer extraction is not always required. For the reduced error correction cycle (not shown), where only 6 stabilizers are mapped out, the probabilities to have a logical X-error is reduced with respect to the overcomplete error correction cycle, while the probability to have a Z-error is increased.

The only logical error on the encoding of $\ket{0_L}$ are X-errors. Since physical CZ gates can only have Z-errors, the bare logical CZ-gate also only gives rise to Z-errors. Especially logical Z-errors on both logical qubits are strongly suppressed, since they can only occur if two physical CZ gates fail on both involved qubits. The logical T-gate also only introduces Z-errors. The physical CZ- and CS-gates anyway only give rise to Z-errors on the data qubits of the logical qubit. On the magic state  X-errors can appear as well as on the auxiliary qubits during the logical CS-gate. However, the latter lead to wrongly applied S-gates on the logical qubit and thus again to Z-type errors, while the former lead with equal probabilities to logical $S_L$ or $S_L^\dagger$ which is equivalent to a noise channel with Kraus operators $I_L$ and $Z_L$.

So far, we considered bare logical gates. However, the resulting noise model is not sufficient to estimate the probability of some quantum algorithm to fail on a logical level. The reason is that the correctable weight-1 errors are not accounted for in our logical noise model. If several gates are performed successively, the weight-1 errors accumulate and can lead to a logical error. The accumulation of errors can be prevented by performing error correction cycles between the gates. In the worst case, there is an error correction cycle after each gate and an upper bound for the logical error rates can be estimated simulating the logical noise model of a so-called extended rectangle including a gadget framed by error correction cycles \cite{Aliferis2006,Chamberland2016}. As the error correction cycle that follows a gate is at the same time the error correction cycle that precedes the next gate, we can subtract the logical error rate of one error correction cycle from the logical error rate of the extended rectangle to estimate the upper bound on the logical error rates. By default, we use the reduced error correction cycle to frame the bare gates. However, for the logical T-gate the preceding cycle uses an overcomplete set of stabilizers. It depends on the algorithm where and how often error correction cycles are performed and the effective noise per gate lies between the error rates of the bare gates and the extended rectangles. The logical noise model including error correction is shown in light colors in \cref{fig:logicalnoise}.

As expected, the logical error rates for the extended rectangles are higher. In particular, we notice, that also the probability to have a Z-error on both logical qubits after a CZ-gate is increased. This is because such errors can also appear if both preceding error correction cycles have an X-Z-error, or one of the preceding error correction cycles has an X-Z-error and in addition one physical CZ-gate fails. A detailed analysis of the logical noise models is given in \cref{app:lognoisemodel}.

In our analysis, we so far assumed that multi-qubit gates have similar error rates as two-qubit gates. However, high-fidelity CCCZ-gates have not yet been demonstrated. As a lower bound for the break even point, we thus derive the break even point for the QEC-cycle and the T-gate with all CCZ- and CCCZ-gates decomposed by single- and two-qubit gates \cite{Heussen2024,Butt2024}. We find that the break-even point of the QEC-cycle is in this case lowered by a factor of 10, while the break even point of the T-gate decreases by a factor 4. For the QEC-cycle lower error rates and thus higher break even points can be achieved using auxiliary qubits to encode the CCCZ-gate (see \cite{Veroni2024} and \cref{sec:implementation}). However, these results stress that the availability of at least high fidelity three-qubit gates is crucial for measurement-free quantum error correction.

\subsection{Comparison with measurement-based schemes}

\begin{figure}[tb]
  \includegraphics[width= \linewidth]{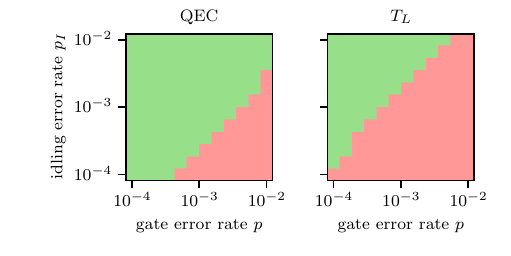}
  \caption{Comparison of the measurement-free QEC-cycle and the T-gate with measurement based schemes. In the green region, the measurement-free schemes have lower logical error rates than the measurement-based version. In the red region measurement-based schemes perform better. The error rates are obtained by summing over the error rates of input states $\ket{0_L}$ and $\ket{+_L}$. }
  \label{fig:comparisonmb}
\end{figure} 

Finally, we compare the performance of our measurement-free schemes with a measurement based approach. Thus, we extend the noise model in \cref{eq:biasednoise} by idling noise and measurement errors. Measurements are performed in the Z-basis and thus measurement errors correspond to a flip $X$ before the measurement takes place. We assume measurement errors to occur with a probability $p_\text{meas}=0.4 p$ \cite{Bluvstein2024}. The dominating error during idling are Z-errors \cite{Bluvstein2022}. Since typical durations of gates are several orders of magnitude smaller than the qubit coherence times, idling errors during gates can be neglected \cite{Levine2019,Evered2023,Pagano2022,Jandura2022}. The noise channel for the idling errors during measurements is given by 
\begin{align}
\mathcal{E}_\text{idle,meas}(\rho)=(1-p_I) \rho + p_I Z \rho Z.
\end{align}

A measurement-based QEC-cycle, that is optimized for the biased noise, is obtained by taking the measurement-free version shown in \cref{fig:main_circuit} and replacing the two gate-based feedbacks with measurements and feed-forward feedback. Similarly, measurement-based magic state preparation optimized for biased noise is given by non-fault tolerant encoding of the magic state, followed by Z-stabilizer extraction and the encoding of a verification flag. Then, all syndrome qubits and the verification-flag are measured and if necessary a second magic state is prepared non-fault tolerantly. The injection is done in the usual way (see for example \cite{Goto2016}). 

In \cref{fig:comparisonmb} we compare the logical error rates of measurement-based and measurement-free schemes for different error rates.

\section{Effects of imperfect noise bias}
\label{sec:imperfectnoisebias}

So far we considered a simplified noise model, where only Z-type errors can happen on two- and multi-qubit gates. In this section we discuss the effects of more general noise models. First, we consider failures on single-qubit gates and afterwards discuss the effect of X- and Y-errors on two- and multi-qubit gates.

\subsection{Faulty single-qubit gates}

\begin{figure}[tb]
  \includegraphics[width= \linewidth]{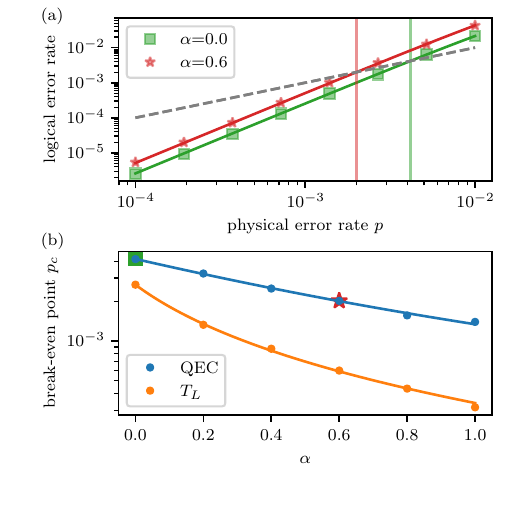}
  \caption{(a) Exemplary determination of the break-even points for the QEC-circuit. (b) Break-even points for the QEC-circuit and the T-gate for different error rates $\alpha=p_1/p$ of the single-qubit gates. The lines correspond to fits of the form of \cref{eq:break_even}.}
  \label{fig:single_noise}
\end{figure}
In addition to the noise model defined in \cref{eq:biasednoise}, we allow X,Y- and Z-errors for the single-qubit gates ($H$, $X$, $T$ and $T^\dagger$) with probabilities $p_1/3$.
The corresponding noise channel is 
\begin{align}
  \mathcal{E}_1(\rho)&=(1-p_1) \rho + p_1/3 \sum_{P \in \{X,Y,Z\} } P\rho P.
\end{align}
Furthermore, we assume the qubit initialization to fail also with probability $p_1$. If the initialization fails a qubit is prepared in $\ket{1}$ instead of $\ket{0}$. This is described by the noise channel
\begin{align}
  \mathcal{E}_I(\rho)&=(1-p_1) \rho + p_1  X\rho X.
\end{align}

We find numerically that all circuits presented in the previous chapter are still fault-tolerant under these conditions. This can be understood in the following way. Typically, fault tolerance is endangered by two- and multi-qubit gates failing simultaneously on data qubits and syndrome qubits and hook errors during syndrome extraction and magic state verification. While simultaneous errors on two qubits can obviously not be induced by faulty single-qubit gates, hook errors might in principle occur. However, the syndrome extraction consists of one H-gate followed by 4 CZ-gates. Thus, if the H-gate or the syndrome initialization fails, the effect on the data qubits is a stabilizer. Any single-qubit error during the magic state verification may result in a logical error on the first magic state, but is then always swapped with the second magic state and the scheme is still fault-tolerant. 

In \cref{fig:single_noise} we show the break-even point for different $\alpha=p_1/p$ for the T-gate and the QEC-cycle. One data point in panel (b) in \cref{fig:single_noise} is obtained in the following way (see also panel (a) in \cref{fig:single_noise}). We simulate the logical error rates for fixed $\alpha$ and different $p$ for initial states $\ket{0_L}$ and $\ket{+_L}$ \footnote{For a gadget with X-, Y-, and Z-errors, the total probability for the gadget to fail is given by the sum of the error rates of X-, Y-, and Z-errors. X-errors are detected with the input state $\ket{0_L}$ and Z-errors with $\ket{+_L}$. However, Y-errors are accounted with both input states, such that the final result gives again an upper bound for the logical error rates.}. We then fit the sum of the error rates for $\ket{0_L}$ and $\ket{+_L}$ with a function of the form $a p^2+b p^3$ and evaluate the break-even point, where the logical error rate is smaller than the physical error rate $p$, with $p_c=(-a+\sqrt{a^2+4b})/(2 b)$. As expected, the break-even point is shifted to smaller error rates with increasing single-qubit failure rates $p_1$. The dependency on $p_1/p$ can be estimated in the following way. The probability to have two errors in the circuit is given by $c_{11} p_1^2 + c_{12} p_1 p + c_{22} p^2$, where $c_{11}$ is the number of possibilities that two single-qubit gates fail, $c_{12}$ the number of pairs of errors on a single and one two-qubit gate and $c_{22}$ the number of errors on two two-qubit gates. The break-even point is then approximately given by 
\begin{align}
p_c=\left(c_{11}\alpha^2+c_{12}\alpha+c_{22}\right)^{-1}.
\label{eq:break_even}
\end{align}

\subsection{Bit-flips on multi-qubit gates}

\begin{figure}[tb]
  \includegraphics[width= \linewidth]{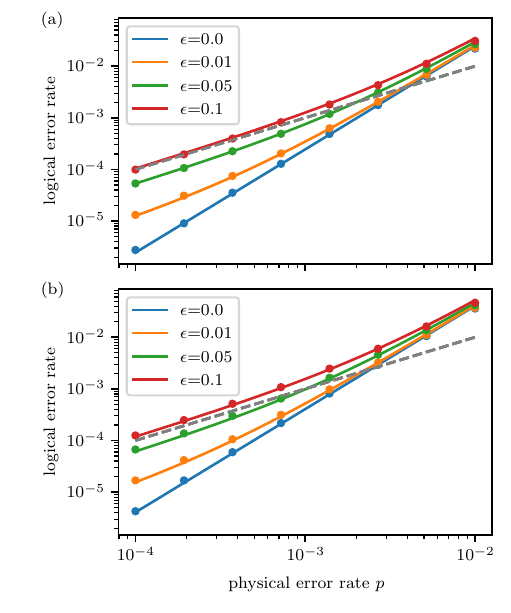}
  \caption{Logical error rates for imperfectly biased noise model as specified in \cref{eq:imperfectbias} for (a) the error correction cycle and (b) the T-gate. The shown error rates are obtained by summing over the error rates of input states $\ket{0_L}$ and $\ket{+_L}$. The lines correspond to the analytical estimates for the logical error rates (see \cref{app:imperfectbiasednoise} for the linear order and \cref{app:lognoisemodel} for the quadratic).}
  \label{fig:imperfectbiasednoise}
\end{figure}

If we allow for X- and Y-errors on two- and multi-qubit gates, the presented QEC-circuit and the T-gate are no longer fault-tolerant. However, if the probabilities for X- and Y-errors errors are significantly smaller than for Z-type errors, our circuits can still suppress the logical error rate. To analyze this quantitatively, we consider the following noise model for two- and multi-qubit gates
\begin{align}
  \tilde{\mathcal{E}}_l(\rho)&=(1-p) \rho + \frac{p(1-\epsilon)}{2^l-1} \sum_{P \in \{I,Z\}^{\otimes l} \backslash I^{\otimes l}} P\rho P \notag \\  &+ \frac{p \epsilon}{2^l(2^l-1)} \sum_{P \in \{I,X,Y,Z\}^{\otimes l} \backslash \{I,Z\}^{\otimes l}} P\rho P,
  \label{eq:imperfectbias}
\end{align}
where $l$ is the number of qubits involved in the gate and $\epsilon$ is the proportion of non-$Z$-type errors. In the limit of $\epsilon=0$ we obtain the perfectly Z-biased noise model. 

In this case single X-errors during syndrome extraction, the verification of the magic state, or feedback can lead to logical errors. However, for small values $\epsilon$ our circuits for the QEC-cycle and the T-gate can still suppress the logical error rate. Furthermore, the logical noise of the T-gate for non-zero $\epsilon$ is still dominated by Z-errors, since the Steane code corrects all types of Pauli errors. In particular, the physical X-errors that destroy the fault tolerance of the logical T-gate can only lead to logical Z-errors. The logical error rates with imperfect noise bias for the QEC-cycle and the T-gate are shown in \cref{fig:imperfectbiasednoise} for different $\epsilon$. In addition, we count the number of errors that destroy fault-tolerance and estimate the critical values for $\epsilon$, above which no break-even point exists (see \cref{app:imperfectbiasednoise}). We find that the analytically found critical value is around $\epsilon_c=0.098$ for the QEC-cycle and $\epsilon_c=0.086$ for the T-gate, which fits the numerical results.

\section{Implementation}
\label{sec:implementation}

\begin{figure}[tb]
  \includegraphics[width= \linewidth]{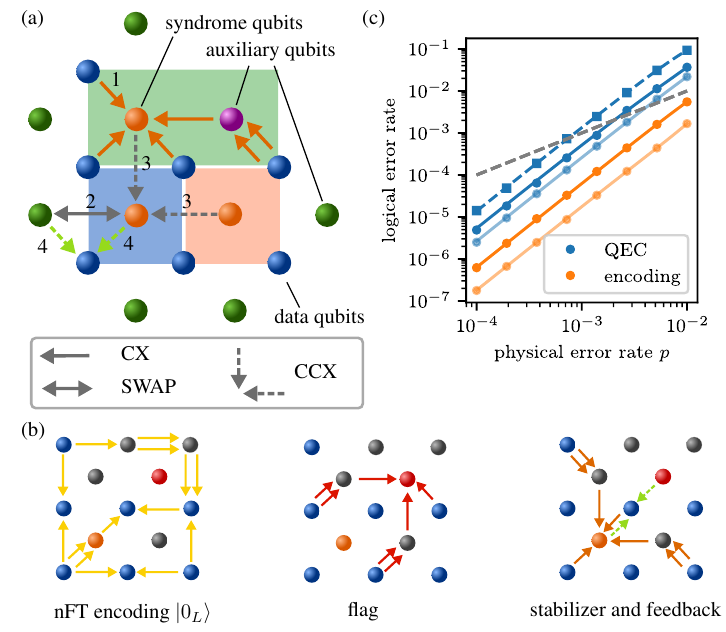}
  \caption{Implementation of an error correction cycle (a) and the encoding of $\ket{0_L}$ (b) on a neutral atom platform with next-nearest neighbor interaction. Straight lines and arrows represent two-qubit gates (CX), and dashed arrows represent three qubit gates (CCX).  Two parallel arrows indicate that after encoding the information into stabilizer or flag qubits the entangling gate is applied again to disentangle the auxiliary qubit. For the error correction cycle this is not shown for step 2 and 3, to keep the figure clear. However, steps 2 and 3 are always reverted. For the error correction cycle step 1 (orange arrows) illustrates the extraction of one stabilizer, while step 2 to 4 illustrate the feedback on one data qubit. For the encoding, we show the full implementation divided into three parts (non-fault-tolerant encoding, flag extraction and stabilizer extraction with feedback). (c) Comparison of the logical error rates of the error correction cycle and the encoding for all-to-all connectivity (light color) and next-nearest neighbor connectivity (dark color). The dashed curve shows the logical error rate for $p_\text{CCZ}=4.2 p$.}
  \label{fig:implementation}
\end{figure}
In this section we exemplarily illustrate efficient implementations of the reduced error correction cycle and the encoding of $\ket{0_L}$ on a neutral atom platform. So far, we assumed all-to-all connectivity. In principle, neutral atom platforms offer this kind of connectivity due to the possibility of shuttling operations. However, tweezer movements are slow and many shuttling operations can be avoided by carefully arranging the qubits in space. Furthermore, the presented protocols include multi-qubit gates like the CCCZ-gate. While such gates are natively available on neutral atom platforms due to the blockade mechanism \cite{Jandura2022,Pelegri2022,Evered2023}, high-fidelity multi-qubit gates have not yet been realized experimentally. Here, we present possible implementations on a two dimensional lattice with next-nearest neighbor interactions, that require at most one shuttling operation and do not require 4-qubit gates. The arrangement of the atoms is shown in \cref{fig:implementation}. The qubits are placed such that most stabilizers can be mapped out without any additional operation. 

For correcting X-errors first all three Z-stabilizers are mapped out into the syndrome qubits (step 1 in \cref{fig:implementation} (a)). While the red and the blue stabilizer can be mapped out without any additional operation, the extraction of the green stabilizer requires an intermediate auxiliary qubit, to perform operations between distant qubits. In particular, potential errors on the data qubit in the bottom right corner of the green stabilizer are copied to the purple auxiliary qubit via a CX-gate \footnote{For simplicity, we use CX- and CCX-gates to describe the implementation of the circuits. However, for the numerical simulation all CX- and CCX-gates are again compiled into H-, CZ-, and CCZ-gates.}. Then, the error-information is transferred to the syndrome qubit via another CX-gate. Finally, the auxiliary qubit and the data qubit are disentangled via an additional CX-gate. For the feedback on one data qubit, the syndrome qubit next to the data qubit has to be swapped with an auxiliary qubit (step 2). Then, the syndrome information of the two distant syndrome qubits is encoded into the auxiliary qubit using a next-nearest neighbor CCX-gate (step 3). Subsequently, a CCX-gate between the auxiliary qubit, the closer syndrome qubit and the data qubit is applied to perform the correction (step 4). Finally, step 3 and 2 are reverted. This is repeated for the feedback on all remaining data qubits always with a new (or reset) auxiliary qubit to ensure fault tolerance. The feedback on the data qubit in the middle does not require the SWAP-step (step 2), here the information of the syndrome qubits of the green and red stabilizer is encoded into the orange auxiliary qubit. For the next round of error correction one either resets all auxiliary qubits (including syndromes) or provides fresh qubits via shuttling. Thus, a full error correction cycle requires at most one shuttling operation, if resets are not available. However, the implementation of error correction with overcomplete stabilizer extraction might require more shuttling operations.

For the encoding we first encode $\ket{0_L}$ non-fault-tolerantly by entangling data qubits. Here, operations between distant qubits can again be performed using intermediate auxiliary qubits. Then, $Z_L$ is mapped to the red flag-qubit. Finally, the stabilizer $X_0X_2X_4X_6$ is mapped to the orange syndrome-qubit and feedback based on the flag and the syndrome qubit is applied to the data qubit in the middle of the Steane-triangle. 

The logical error rates for next-nearest neighbor connectivity (dark color) and all-to-all connectivity with 4-qubit gates (light color) are compared in \cref{fig:implementation} (c). For the QEC-cycle, we in addition show the logical error rates for three-qubit gate infidelities of $p_\text{CCZ}=4.2 p$ \cite{Evered2023}, where $p$ is the two-qubit gate infidelity. For the encoding we obtain similar logical error rates for $p_\text{CCZ}=4.2 p$ and $p_\text{CCZ}=p$. The logical error rates are higher than for all-to-all connectivity, due to the additional gates required for the next-nearest neighbor connectivity. However, the break-even points are still in an experimentally realistic regime.

\section{Conclusions and outlook}
\label{sec:conclusions}

In this work we construct measurement-free error correction protocols and a universal gate set for setups, where the noise is dominated by phase-errors on two- and multi-qubit gates. The noise bias enables us to reduce the qubit- and gate- overhead compared to previous measurement-free proposals \cite{Heussen2024,Veroni2024,Butt2024,Veroni2024_2} and thus to push the break-even point into a regime that is feasible with state-of-the-art neutral-atom experiments. We analyze the logical noise model of the logical qubits as well as the impact of imperfections in the noise bias. For an asymmetric depolarizing noise model we find that the presented schemes can be used to suppress the error rate if more then $90\%$ of the errors in two- and multi-qubit gates are Z-errors. 

In our work we focus on the Steane code, however, we expect that the discussed protocols can be extended to other low-distance CSS-codes. Here, the detailed discussion of the circuit construction with biased noise in \cref{sec:logicalqubits} might serve as instructions to reduce the measurement-free QEC-circuit for other codes \cite{Heussen2024,Veroni2024} and to introduce generalized measurement-free magic state injection schemes.

Our protocols allow to correct up to one Z- or X-error on the logical qubits. While this is not sufficient for suppressing errors arbitrarily well, our logical qubits might be used to push the logical error rate below the threshold of a fully fault-tolerant and scalable error-correcting code.  The idea is then to use our logical qubits as data qubits for higher level error correction protocols. In particular, we want to point out that these protocols include measurement-based error correction schemes, even if measurements are slow. The reason is, that the Steane code can be used to correct any single-qubit Pauli-error, including errors that might occur while waiting for measurement results. Thus, the probability to have logical idling errors can also be suppressed. In conclusion, logical qubits encoded in the Steane code together with our platform-specific protocols might be a good starting point for the implementation of arbitrary scalable and universal error-correcting codes. 

\section{Acknowledgements}
We acknowledge funding by the Federal Ministry of Education and Research (BMBF) project MUNIQC-ATOMS (Grant No. 13N16070).
H.P.B., S.W., S.H.Q. and K.B. additionally acknowledge funding from the BMBF under the grant QRydDemo, and from the Horizon Europe program HORIZON-CL4-2021-DIGITAL- EMERGING-01-30 via the project 101070144 (EuRyQa).
F.B., D.L. and M.M. additionally acknowledge support from the German Research Foundation (DFG) under Germany’s Excellence Strategy ‘Cluster of Excellence Matter and Light for Quantum Computing (ML4Q) EXC 2004/1’ 390534769,
the BMBF via the VDI within the project IQuAn,
the ERC Starting Grant QNets through Grant No. 804247, 
the US Army Research Office through Grant Number W911NF-21-1-0007, 
the Intelligence Advanced Research Projects Activity (IARPA) under the Entangled Logical Qubits program through Cooperative Agreement Number W911NF-23-2-0216, 
the European Union (EU) Horizon Europe research and innovation program under Grant Agreement No. 101114305 (“MILLENION-SGA1” EU Project),
and the Munich Quantum Valley (K-8), which is supported by the Bavarian state government with funds from the Hightech Agenda Bayern Plus.

\nocite{DARUS-5455_2025}

\appendix

\section{Details of the biased noise model}
\label{app:detailsnoisemodel}

The ability to have a strong noise bias for two-qubit gates on Rydberg platforms has been discussed in several theory proposals \cite{Cong2022,Sahay2023}. The main idea is that leakage errors can be converted into Z-type errors. In the following we discuss this in more detail.

The fidelities of two- and multi-qubit gates are fundamentally limited by the finite lifetime of the Rydberg state. The ability to entangle qubits and thus the need of interactions between qubits require temporary population of the Rydberg state. However, Rydberg states suffer from decay and thus imprint decay channels on the qubit states. The possible decay processes depend on the encoding of the qubits and the choice of the Rydberg state. For the following discussion we need to distinguish three scenarios. 

\begin{itemize}
\item Decay to the qubit state $\ket{0}$.
\item Decay to the qubit state $\ket{1}$.
\item Decay out of the qubit subspace.
\end{itemize}

In typical implementations of CZ-gates, CP-gates or CCCZ-gates, only the qubit state $\ket{1}$ has to be coupled to the Rydberg state \cite{Pagano2022,Jandura2022}. Thus, for such two- or multi-qubit gates the qubit state $\ket{1}$ decays effectively via the channels discussed above, while the $\ket{0}$-state is not affected. Consequently, decay to the qubit state $\ket{1}$ leads to phase type errors.
In case the atom left the qubit subspace it can be reinitialized in the qubit subspace using for example optical pumping \cite{Cong2022} or SWAP-type gate operations \cite{Chow2025,Perrin2024}. The state, in which the qubit is reinitialized determines the final effective noise. If it is initialized in a maximally mixed state the noise is described by a depolarizing noise model. However, if the qubit is reinitialized in the state $\ket{1}$, no bit-flips occured and the only errors that can appear are phase errors, as the qubit now effectively decayed from the state $\ket{1}$ to the state $\ket{1}$. 

Decay to the $\ket{0}$-state leads inevitably to X-type errors and also in the reinitialization imperfections might lead to population of the $\ket{0}$-state and thus to X-errors. However, for the example of $^{171}$Yb the proportion of errors that are converted to Z-errors is expected to be around $98 \%$ \cite{Sahay2023}.

\section{Numerical simulations}
\label{app:numerics}

For the biased noise model we verify fault-tolerance of all circuits by simulating all possible faulty circuits, where one gate fails. 
To determine  the logical noise model and the logical error rates (\cref{fig:logicalnoise} and \cref{fig:implementation}), we then simulate $10^4$ circuits with minimum 2 randomly placed faults. 

For the imperfect biased noise (\cref{fig:imperfectbiasednoise}) we know that single faults can already lead to logical errors. Thus, we simulate $10^6$ circuits with minimum 1 randomly placed fault. 

All simulations are performed using exact state vector simulations in qiskit \cite{Javadi-Abhari2024}.

\section{Detailed analysis of the logical noise model}
\label{app:lognoisemodel}

In addition to the numerical simulations we estimate the logical noise model analytically. Furthermore, we show the noise model of the CZ-, T-, H, and S-gate surrounded by QEC-cycles.

\subsection{Error correction cycle}

\textit{Reduced error correction cycle }- Initially, we calculate the error-rates on the physical data qubits based on the number of faulty-gates they are involved in. Then, we count the number of possibilities to place two of those single-qubit errors and estimate the logical error rates. We consider the biased noise-model defined in \cref{eq:biasednoise}. Technically, there are only Z-errors, however, since the correction of Z-errors is done by framing the circuit for correcting X-errors with H-gates, Z-errors can be transformed into X-errors.

During the simplified error correction cycle, a single X- or Z-error on the data qubits happens if a CZ-gate during the syndrome extraction fails on a data qubit (the probability for this is $2/3 p$ for the qubits in the corner, $2 \times 2/3 p $ for the sides and $3 \times 2/3 p$ for the center of the triangle) or if the feedback gate fails (the probability for this is $8/15$ for all qubits). In addition, an error on the qubits in the corner happens if a syndrome qubit is flipped during the stabilizer extraction ($ 4\times 2/3$).  Thus, the total probability to have a single X-error is $p_c=58/15p$ for the corner, $p_s=28/15 p$ for the sides and $p_m=38/15 p$ for the center. There are 3 possibilities to place two errors on corners or sides, 3 to place one error on a corner and one on a side and 3 each to place one error in the middle and one on a side or a corner. In leading order one obtains then a probability of $p_{X,rQEC}=142.5 p^2$ to have a logical X-error. 

The probability to have a logical Z-error is smaller, since Z-errors that happen during the correction of X-errors might be corrected in the second half of the error correction cycle. Thus we have to subtract all combinations of Z-errors that come from the first stabilizer extractions and Z-errors during the second feedback and obtain $p_{Z,rQEC}=120.1 p^2$. 

However, some of the discussed error combinations lead to X- and Z-errors and thus to logical Y-errors. Logical Y-errors can only happen if there are at least two X- and two Z-errors on the data qubits. This is only the case if one CZ-gate fails on both qubits during the extraction of the Z-stabilizers and one during the extraction of the X-stabilizers. The probability for one CZ-gate to fail on both qubits is $p/3$. There are 12 CZ-gates during the extraction of the Z-stabilizers and thus 12 possibilities for a single fail, so in total 144 possibilities for one fail during X-stabilizer extraction and one during Z-stabilizer extraction. From this we have to subtract all configurations where the two Z- or X-errors cancel out. Errors can cancel out if they happen on the corner of the Steane-triangle, so there are 3 options to have an X-or Z-error on the corner and 4 for the corresponding syndrome to be flipped.  Thus, the probability for a logical Y-error is $(144-2\cdot 3\cdot4) \cdot (p/3)^2=13.3 p^2$. Subtracting this from the previously estimated probability for a logical X- or Z-error we obtain the following probabilities for a logical X-,Y- or Z-error
\begin{align}
p_{X,rQEC}&=142.5 p^2-13.3 p^2=129.2 p^2, \notag \\
p_{Y,rQEC}&=13.3 p^2, \notag \\
p_{Z,rQEC}&=120.1 p^2-13.3 p^2=106.8 p^2.
\end{align}

\textit{Error correction cycle} - In the error correction with overcomplete stabilizer extraction the probability to have an X-error due to syndrome extraction on the corner and in the middle is increased ($4/3$ for the corner and $8/3$ for the middle). The probability to have a Z-error is reduced, since errors on syndromes only lead to logical errors if there is already a single Z-error in the beginning of the Z-error correction. It is thus convenient to initially only take into account the probability to have Z-errors during the syndrome extraction. The probabilities for single Z-errors are then $p_c=2/3$, $p_s=4/3$ and $p_m=2$ and the probability for a logical Z-error is then $26.7 p^2$. In addition a logical Z-error can appear if two CCCZ-gates in the feedback fail. The probability for this is $21 (8/15)^2 p^2$. And finally if there is a Z-error on the side (in the middle) of the triangle there are two (three) possibilities, where an additional error on the syndromes lead to a logical error. This gives additional $37.3 p^2$. In total, we thus obtain $70 p^2$. Again, we did not yet take into account Y-errors. Logical Y-errors appear only if during the first and during the second round of stabilizer extraction a CZ-gate fails on both qubits. Since with the overcomplete stabilizer extraction logical Z-errors are less likely, we initially determine the probability that two failing CZ-gates give a logical Z-error. As discussed before this can only happen if the first Z-error is on the side or in the middle of the triangle, which gives a probability of $37.3/4 p^2$. (Here the factor of $1/4$ comes in, since this time we only allow for errors on both qubits, that happen with a probability of $1/3p$, while previously we also allowed for errors on one qubit ($2/3p$)). From this we have to subtract the probability to have two X-errors on the same corner, which is $24/9 p^2$. Finally we obtain
\begin{align}
p_{X,QEC}&=181.7 p^2, \notag \\
p_{Y,QEC}&=6.7 p^2, \notag \\
p_{Z,QEC}&=63.3 p^2.
\end{align} 

\subsection{Encoding}

For the encoding of logical $\ket{0_L}$ we first notice that only X-errors have a non-trivial effect on the encoded logical state. Since physical CZ- and CCZ-gates only give rise to Z errors, the only data qubits that can have an X error are the qubits 0,1,3 and 5 where a H-gate is performed after a CZ- or CCZ-gate. The probability for an X-error on the qubit 0 is $2 \times 2/3 p +4/7 p$ and for 1,3, and 5 the probability is $2 \times 2/3 p$. There are 3 possibilities to place 2 errors on 1,3 and 5 and 2 possibilities to place one error on 1 or 5 and one on 0. Errors on 3 and 0 cancel out. Thus, the probabilities to have a logical X error after the bare non-fault-tolerant encoding is approximately $10.6 p^2$. Furthermore, a logical X-error appears if there happens an error on qubit 3 and in addition the verification flag or the syndrome is flipped. The probability for this is $6.2 p^2$ thus the probability to have a logical error in the encoding is
\begin{align}
p_{X,Enc}&=16.8 p^2.
\end{align}

\subsection{CZ-gate}

For the logical CZ-gate a logical error on only one logical qubit happens if two gates fail either on the same qubit or one gate fails on one qubit and one gate on both. For the first case there are 21 possibilities with probability $p^2/9$, while for the last case there are 42 possibilities with probabilities $p^2/9$. The total probability for a logical $ZI$ error or $IZ$ error is thus 
\begin{align}
p_{ZI,CZ}=63/9 p^2=7 p^2. 
\end{align}
The probability for a logical $Z$ error on both logical qubits is smaller, since the only possibility is that 2 physical CZ-gates fail on both qubits. Thus, there are 21 possibilities with probabilities $p^2/9$ and the probability for a logical $ZZ$ error is 
\begin{align}
  p_{ZZ,CZ}=7/3 p^2. 
\end{align}

\subsection{T-gate} 

Logical errors during the T-gate with magic state injection can appear in the following way. Either errors happen directly on the logical state if one of the CZ- or CS-gates fail or errors that happen on the magic qubit imply errors on the logical state. 

We start with the effects that errors on the magic qubit have on the logical qubit. Z-errors on the magic state flow to the logical qubit via the logical CX-operation in the beginning of the injection and thus directly translate into Z-errors on the logical qubit. Logical X- and Y-errors on the magic state (also during the logical CX-gate in the injection) imply a wrongly applied logical $S_L$ if the magic qubit is in $\ket{0_L}$ and a logical $S^\dagger_L$ if the magic qubit is in $\ket{1_L}$. After tracing out the magic qubit this leads to a logical noise channel with Kraus operators I$_L/\sqrt{2}$ and Z$_L/\sqrt{2}$ on the logical qubit. Flips on the two flags during the logical CS-gate with subsequent reset have the same effect as well as a single X-error on the magic state data qubits 4,5 or 6 in  combination with an X-error on one of those flags. 

Taking into account the previous considerations and the fact that the CZ- and the CS-gates give only rise to Z-errors, the logical T-gate can only have Z-errors. The probability for a logical Z error is given by half the probability to have a logical X-error on the magic state and the probability to have a weight-2 Z-error during the magic state preparation and injection. In the following we derive both probabilities.

Single X-errors on the magic state can appear after the CX gate with probability $2/3 p$, after the CSWAP gate with probability $2/3 p$ and in the verification with probability $1/3 p$. For the verification we do not count errors where the gate fails on both qubits, since this results in a raised flag. However, if two gates during the verification fail on both qubits the flag is flipped twice and the logical qubit has a logical X-error. Thus, we have to add additional $21/9 p^2$. The total probability to have a single X-error on the magic state is thus $(2/3+4/3+1/3)p=7/3 p$. Finally, to flip one of the flags in the logical CS-gate with subsequent reset either an error has to happen on the syndrome qubits for $Z_0 Z_1Z_2Z_3$ and $Z_0 Z_1 Z_5 Z_6$ or during the encoding of the flag. The probability for such a flip is thus $(4\cdot2\cdot 2/3+8/15) p=98/15$.
In summary, the probability for a weight-2 X- or Y-error on the magic qubit is then 
\begin{align}
p_{X,M}&=\left(21 \left(\frac{7}{3}\right)^2+\left(\frac{98}{15}\right)^2+2\cdot 3 \frac{7}{3}\frac{98}{15}+\frac{21}{9}\right)p^2\notag \\&\approx 251 p^2.
\end{align}

Single Z-errors on the logical qubit can appear after the CZ- and CS-gate each with a probability of $2/3 p$. Another possibility to have Z-type errors on the logical qubits is an accidental flip of one control qubit that controls the CS-gate. The probability for this is $0.5(3 \cdot 2/3+4/7)p$, where  factor of $0.5$ comes in because a wrongly applied S is only half an Z-error. Furthermore, logical Z-errors on the magic state appear during the CSWAP-type feedback with a probability of $4/7p$ and in the verification with a probability of $1/2 \cdot 1/3 p$. Again, for the verification we only have to count errors that happen on both qubits, if two such errors happen, since otherwise the flag is raised. However, weight-2 errors in the verification  give also always rise to logical X-error and are thus already taken into account. Similarly, Z-errors on both qubits during the CZ-gate in the injection lead to logical X-errors. Thus, we have to subtract $(1+1/4) \cdot 21/9p^2$ from the final end-result. Single Z-error during the syndrome extraction in the magic state preparation raises the flag and thus do not matter. However, a logical error during the syndrome extraction can appear with probability $26/3$. The probability to have a single Z-error on one data qubit is thus $(2/3+2/3+(3 \cdot 2/3+4/7)/2+4/7+1/6)p\approx 3.36p$ and the probability to have a weight-2 Z-error
\begin{align}
  p_{Z,a}&=\left(21\cdot 3.36^2 -1.25\cdot \frac{21}{9}+\frac{26}{3} \right)p^2\notag \\&\approx 242 p^2.
\end{align}

Last but not least logical Z-errors on the magic state can appear, if the second non-fault-tolerant magic state encoding fails and in addition the flag for doing the SWAP is raised. There are only two error positions in the non-fault-tolerant encoding that can lead to a logical Z-error and the probability is thus $4/3 p$. The flag is raised if there is a single $Z$ error during the first encoding ($11p$), if there is a single Z-error during the Z-stabilizer extraction on the data qubits ($12\cdot 1/3 p$), if there is an error on the syndrome qubits ($12 \cdot 2/3 p$), an error on the ancilla in the verification ($7\cdot 2/3 p$) or during the final flag encoding ($3\cdot8/15 p$). Errors on data qubits during the X-syndrome extraction count only half, since they raise the flag in only half of the runs. The total probability for this scenario is then 

\begin{align}
  p_{Z,b}&=\left(11+ \frac{12}{3}+\frac{24}{3}+\frac{14}{3}+\frac{24}{15}  \right) \frac{4}{3}p^2\notag \\&\approx 39 p^2.
\end{align}

In summary a logical Z-error appears after a logical T-gate with a probability of 
\begin{align}
p_{Z,T}=(p_{X,M}/2+p_{Z,a}+p_{Z,b}) p^2 \approx 406 p^2, 
\end{align}
where the factor $1/2$ comes in, since a logical X-error on the magic state implies only half of the time a logical Z-error on the logical qubit.

\subsection{Extended rectangles} 

\begin{figure}[tb]
  \includegraphics[width= \linewidth]{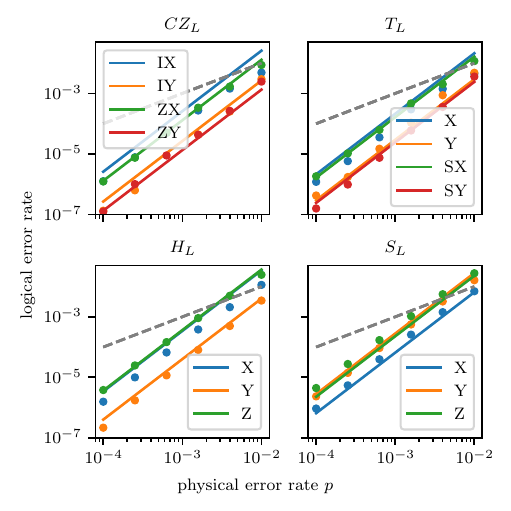}
  \caption{Characterization of logical noise model for the logical CZ-gate, the logical T-gate, the logical H-gate and the logical S-gate surrounded by QEC. The points show the numerically simulated data and lines are the analytical estimates for the logical error rates. For the CZ- and T-gate Z-type errors are shown in \cref{fig:logicalnoise}. Thus, we do not show them again.}
  \label{fig:fulllogicalnoise}
\end{figure}

To determine the error probabilities of the extended rectangle, we use the coefficients determined above and combine their square-roots to estimate the error-rates of the extended rectangles. Here, the idea is that the number of possibilities to place 2 errors in a circuit is roughly given by the number of possibilities to place 1 error multiplied with itself. We first estimate analytical values for the curves shown in \cref{fig:logicalnoise} and afterwards discuss logical X- and Y-type errors that occur for the logical CZ-, T-, H, and S-gate surrounded by QEC-cycles, but do not originate from the bare gates.

For the extended encoding we then obtain
\begin{align}
p_{X,extEnc}&=(\sqrt{p_{X,Enc}}+\sqrt{p_{X,rQEC}})^2\approx 239.5 p^2. 
\end{align}

For the extended CZ-gate we obtain
\begin{align}
p_{ZI,extCZ}&=(2 \sqrt{p_{Z,rQEC}}+\sqrt{p_{ZI,CZ}})^2 \notag\\ &+(\sqrt{p_{X,rQEC}}+\sqrt{p_{Y,rQEC}})\notag \\ &\times(\sqrt{p_{Z,rQEC}}+\sqrt{p_{Y,rQEC}}+\sqrt{p_{ZI,CZ}})\notag \\&-p_{Z,rQEC} \notag \\ &\approx 936.144 p^2, \notag \\
p_{ZZ,extCZ}&=(\sqrt{p_{Y,rQEC}}+\sqrt{p_{ZZ,CZ}})^2 \approx 26.8,
\end{align}
where we also take into account that X-errors on data qubits lead to Z-errors on the data qubits of the other logical qubit after the CZ-gate.

For the extended T-gate we use
\begin{align}
p_{Z,extT}&=(\sqrt{p_{Z,rQEC}}+\sqrt{p_{Z,QEC}}+\sqrt{p_{Z,T}})^2-p_{Z,rQEC}\notag \\ &\approx 1372.28 p^2.
\end{align}

In \cref{fig:fulllogicalnoise} we show the probability for X- and Y-type errors to occur in the extended rectangle CZ- and T-gate and the logical noise model of the extended logical H- and S-gate.

For the CZ-gate a logical X-error in the first QEC-round leads to a logical Z-error on the other logical qubit. In \cref{fig:fulllogicalnoise} we refer to this as ZX, similarly a logical Y-error before the bare logical CZ-gate leads to ZY. Thus, we find
\begin{align}
p_{ZX,extCZ}&=p_{X,rQEC} \notag \\
p_{ZY,extCZ}&=p_{Y,rQEC}.
\end{align}
A single physical X-error before the CZ-gate will not affect the logical information of the other qubit and thus can only give IX-type errors, if in addition an X-error happens during the second round of error correction. Similar arguments hold for IY. We thus estimate the probability for IX- and IY-errors as  
\begin{align}
p_{IX,extCZ}&=2 p_{X,rQEC} \notag \\
p_{IY,extCZ}&=2 p_{Y,rQEC}, 
\end{align}
where the factor two accounts for the fact that the number of possibilities to place two errors in two different parts of the circuit is twice the number of possibilities one has when placing them in the same part, since in the first case they are distinguishable while in the second case they are not. However, our analytical approximations of $p_{IX,extCZ}$ and $p_{IY,extCZ}$ seem to overestimate the logical error rates (see \cref{fig:fulllogicalnoise}). The reason is that some of the single physical X-errors that occur during the first round of QEC might be corrected in the second round of QEC, before weight two-errors can accumulate.

As expected, the T-gate transforms logical X-errors into SX-type errors and logical Y-errors into SY-type errors. Naively on would thus assume that error rates for logical SX-errors (respectively SY-errors) is given by the probability to have an X-error (Y-error) in the QEC-cycle. However, our realization of the logical T-gate transforms combinations with one physical X-error and one physical Y-error half of the time into a logical SX-error and half of the time into a logcial SY-error. The logical error rates for SX- and SY-errors can be approximated as
\begin{align}
p_{SX,extT}&=p_{X,QEC}-\sqrt{p_{X,QEC}p_{Y,QEC}}/2\notag  \\ &\approx 164.305 p^2 \notag \\
p_{SY,extT}&=p_{Y,QEC}+\sqrt{p_{X,QEC}p_{Y,QEC}}/2\notag \\ &\approx 24.069 p^2.
\end{align}
X- and Y-errors can occur if two X- and Y-type errors occurs in the second error correction round or if one appears in the first and one in the second. Since we always substract the probability of one error correction cycle to fail, we only have to count the combinations, where one error happend in the first QEC-cycle and one in the second. The T-gate transform physical X-errors into SX-errors which corresponds to a equal weight superposition of an X- and an Y-error and similarly for incoming physical Y-errors. Logical Y-errors occur only if the final errors are two Y-error and the probabilities can thus be estimated as 
\begin{align}
p_{X,extT}&=(\sqrt{p_{X,QEC}}+\sqrt{p_{Y,QEC}})\sqrt{p_{X,rQEC}}\notag \\&+(\sqrt{p_{X,QEC}}+\sqrt{p_{Y,QEC}})\sqrt{p_{Y,rQEC}}/2\notag  \\ &\approx 211.874 p^2 \notag \\
p_{Y,extT}&=(\sqrt{p_{X,QEC}}+\sqrt{p_{Y,QEC}})\sqrt{p_{Y,rQEC}}/2\notag  \\ &\approx 29.3244 p^2  
\end{align}

The H-gate transforms X-errors into Z-errors and vice versa. Thus, the probability to have logical X-, Y- and Z-errors in the extended logical H-gate can be estimated as 
\begin{align}
p_{X,extH}&=(\sqrt{p_{Z,rQEC}}+\sqrt{p_{X,rQEC}})^2-p_{X,rQEC}\notag  \\ &\approx 341.653 p^2 \notag \\
p_{Y,extH}&=(\sqrt{p_{Y,rQEC}}+\sqrt{p_{Y,rQEC}})^2-p_{Y,rQEC}\notag \\ &\approx 39.999 p^2 \notag \\
p_{Z,extH}&=(\sqrt{p_{X,rQEC}}+\sqrt{p_{Z,rQEC}})^2-p_{Z,rQEC}\notag \\ &\approx 364.053 p^2.
\end{align}
Again, our analytical approximation slightly overestimates the logical error rates for logical X-errors, since we do not account for possible corrections of single physical errors in the second round of QEC.

The S-gate transforms X-errors into Y-errors and vice versa. Thus, the probability to have logical X-, Y- and Z-errors in the extended logical H-gate can be estimated as 
\begin{align}
p_{X,extS}&=(\sqrt{p_{Y,QEC}}+\sqrt{p_{X,rQEC}})^2-p_{X,rQEC}\notag  \\ &\approx 65.3575 p^2 \notag \\
p_{Y,extS}&=(\sqrt{p_{X,QEC}}+\sqrt{p_{Y,rQEC}})^2-p_{Y,rQEC}\notag \\ &\approx 280.149 p^2 \notag \\
p_{Z,extS}&=(\sqrt{p_{Z,QEC}}+\sqrt{p_{Z,rQEC}})^2\notag -p_{Z,rQEC}\notag \\ &\approx 227.723 p^2.
\end{align}

\section{Analytical arguments for imperfect biased noise}
\label{app:imperfectbiasednoise}

Here, we estimate the critical values for the noise bias $\epsilon$.

\subsection{Error correction cycle}

\textit{Reduced error correction cycle} - In the reduced error correction cycle there are 4 possible scenarios, where X or Y errors on one physical gate lead to a logical X- or Z-error. In the following we discuss the possibilities to cause logical X-errors. Due to symmetry reasons, the probabilities to have logical Z-errors are the same. 

The first scenario are flips of the syndrome qubits during the stabilizer extraction that lead to hook errors. For each stabilizer extraction block the critical CZ-gates are the second and the third. (the second if there is no Z-error on the involved data qubit and the third if there is a Z-error on the involved data qubit). In both cases the probability for one CZ-gate to have such an error is approximately $1/3 \epsilon p$. In total there are 3 stabilizer extraction blocks, where hook errors imply logical X-errors and thus the probability for a logical X-error due to hook errors is $2 \epsilon p$. 

The second scenario are X-errors on the data qubits (but not the syndromes) during the extraction of the Z-stabilizers that lead to wrong syndromes and thus to wrong corrections. Errors during the extraction of the last stabilizer do not matter. During the extraction of the second stabilizer there are only two qubits, that are also involved in the extraction of the last stabilizer and in the first stabilizer there are three qubits (all except the corner) that are also involved in the extraction of the other stabilizers. The probability for one CZ-gate to have an X- or Y-error on the data qubit is $1/4 \epsilon p$ and thus the probability for a logical error due to the second scenario is $5/4 \epsilon p$. 

The third scenario are X-errors on the data qubits in combination with flips of the syndromes that can appear during the stabilizer extraction. Here, during the first extraction errors do not matter, while during the extraction of the second stabilizer there are two problematic CZ-gates and during the extraction of the third stabilizer there are 3. Thus, the probability for a logical error due to the third scenario is $5/4 \epsilon p$. 

The last scenario are flips on the syndrome qubits during the feedback, that lead to wrong corrections. If the gate that flips the syndromes at the same time also fails on the involved data qubit this results in a logical error. If the first feedback gate fails there are 6 possibilities for syndrome flips that imply a wrong correction. For the second feedback gate there are 5 possibilities, for the third feedback gate four and so on. The total probability for a logical X-error due to the third scenario is then $21/15 \epsilon p$. In total the probability for a logical X-error as well as a logical Z-error is thus $5.9 \epsilon p+\mathcal{O}(p^2)$ and the probability to have any logical error is $11.8 \epsilon p +\mathcal{O}(p^2)$. Thus, the critical value for $\epsilon$ is $\epsilon_c \approx 0.084$. 

\textit{Error correction cycle} - For the error correction with overcomplete stabilizer extraction the probability for hook errors is $4 \cdot 2/3 \epsilon p$. However, there are only two problematic X-errors on the data qubits during the extraction of the Z-stabilizers and thus the probability for a logical error during the second and third scenario is each $\epsilon p/2$. Furthermore, during the feedback there are only 3 problematic errors during the first CCCZ-gate, 2 during the second and third and one during the fourth. The probability for a logical error during the feedback is thus $3/5\epsilon p$. In total the probability for a logical error  is then $(5.9+4.266) \epsilon p+\mathcal{O}(p^2) \approx 10.17 \epsilon p+\mathcal{O}(p^2)$ and the critical value for $\epsilon$ is $\epsilon_c \approx 0.098$.

\subsection{T-gate}  

For the T gate with magic state injection we initially consider the magic state preparation. First, we notice that now weight-2 Z-errors, that are not detected by the verification, can appear if in the non-fault-tolerant encoding of the magic state a physical CZ-gate has an X-Z-error. The probability for this is $9/12$.  Next, we consider hook errors during the extraction of the Z-stabilizers and notice that Y-errors on the syndrome trigger the flag as well, so the only problematic errors are X-errors on syndromes, which appear with a probability of $2/6 \epsilon p$. Thus, the contribution is $3\times 2/6 \epsilon p = \epsilon p$. Hook errors in the verification lead to logical X-errors, which as discussed in \cref{app:lognoisemodel} gives rise to logical Z-errors in the injection with a probability of 50\%. Errors during the last gate in the verification do not matter and again, only X errors have to be counted, as Y-errors also raise the flag. In summary, the contribution from the verification is thus $0.5 \times 6 \times 2/6 \epsilon p =\epsilon p$. In the CSWAP-type feedback errors on the control between the second and fifth swap lead to X-errors on the data qubits if the two qubits participating in the swap are in different states, which is roughly the case half of the time.  The probability for one CSWAP (disassembled into three CCZ and H) to flip the control qubit is $3\times 4/7 \epsilon p$ and the total contribution from the CSWAP-type feedback is $0.25\times 5\times3\times 4/7 \epsilon p \approx 2 \epsilon p$, where the second factor $1/2$ stems from the fact that logical X-errors on the magic state imply only half of the time a logical Z-error on the data qubits.

During the logical CS-gate with subsequent reset logical errors occur if the qubit 4,5 or 6 are flipped during the syndrome extraction or the extraction of Z$_L$. Both lead again to an S-type logical error on the logical qubit and thus lead to a logical error only half of the time.  In each round of syndrome extraction there are 5 gates where such a flip can happen. During the second round the probability for such a flip is $2/3 \epsilon p$ and during the first round it is  
$1/2 \epsilon p$, since here the combination Y-error on data qubit and Z-error on the syndrome qubit is still correctable via the two flags. The total contribution from the syndrome extraction is thus $0.5 \times 4 \times( 2/3+1/2) \epsilon p  = 2.34 \epsilon p$. 

A flip during the last round of Z$_L$ extraction does not matter, a flip in the round before gives a weight-1 and thus correctable error and a flip in the 5th round gives a weight-2 S-type error and thus only 1/4 logical error. Flips in the rounds 1 to 4 lead to a logical S-error. The contribution from logical Z extraction is thus $0.5\times 4 \times 3 \times 2/3 \epsilon p+ 1/4\times 3 \times 2/3 \epsilon p =4.5 \epsilon p$.

The total probability for a logical error during the T-gate is then $11.5 \epsilon p+\mathcal{O}(p^2)$. Thus, the estimated critical value for $\epsilon$ is $\epsilon_c \approx 0.86$.

\section{Resources and circuits}
\label{app:circuits}

\begin{figure}[p]
  \includegraphics[width= \linewidth]{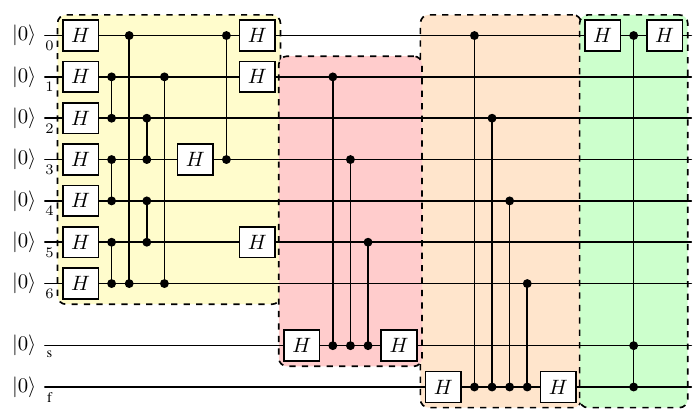}
  \caption{Circuit for encoding logical zero on the $[[7,1,3]]$ Steane code \cite{Heussen2024}.}
  \label{fig:full_encoding_logical_zero}
\end{figure}

\begin{figure*}[p]
  \includegraphics[width= \linewidth]{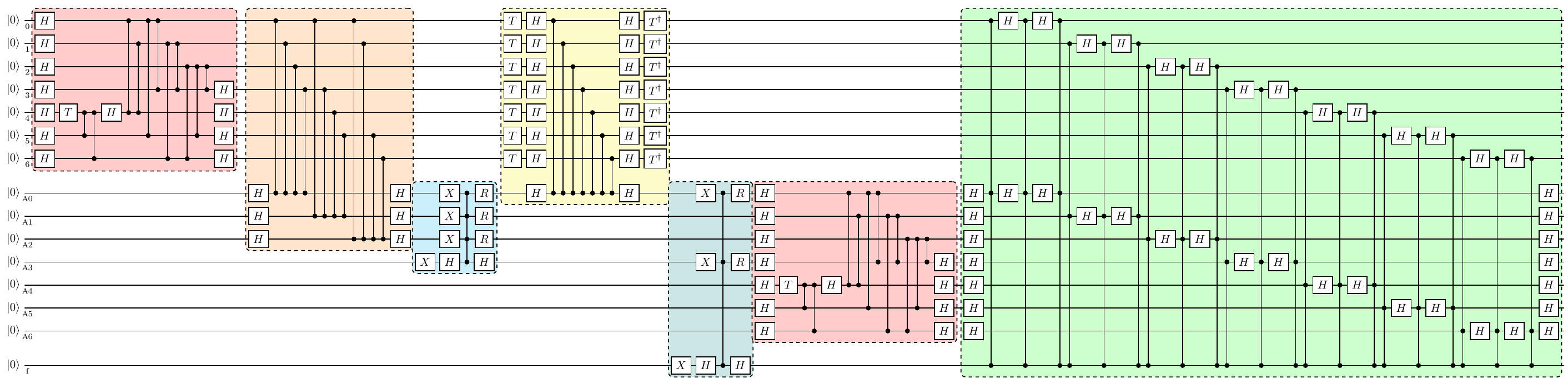}
  \caption{Circuit for magic state encoding. The colors are chosen to fit \cref{fig:main_circuit}.}
  \label{fig:full_magic_encoding}
\end{figure*}

\begin{figure*}[p]
  \includegraphics[width= \linewidth]{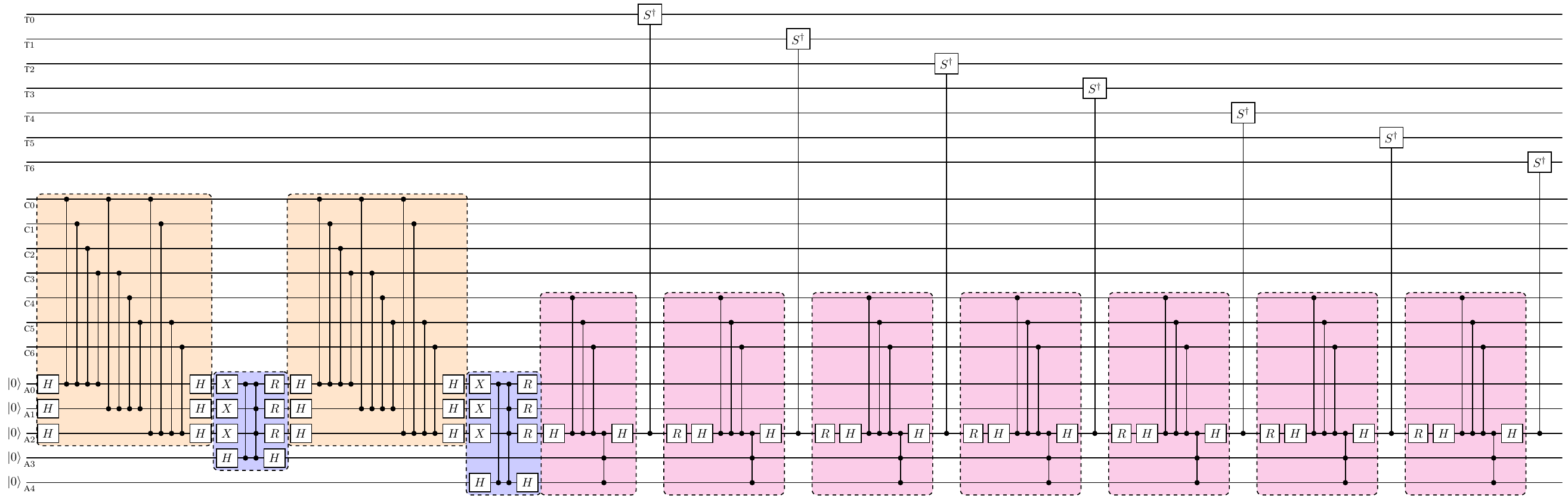}
  \caption{Circuit for the logical CS-gate with subsequent reset on the control. The colors are chosen to fit \cref{fig:main_circuit}}
  \label{fig:full_cs}
\end{figure*}

\cref{fig:full_encoding_logical_zero} shows the full circuit for encoding $\ket{0_L}$ on the Steane code, \cref{fig:full_magic_encoding} shows the full circuit for magic state encoding, and  \cref{fig:full_cs} shows the full circuit for the logical CS-gate.

\begin{table}
  \begin{tabular}{ |l|c c c c c| } 
   \hline
                      & CZ & CCZ & CCCZ & CP &qubits \\ \hline
    reduced QEC        & 24 & 0 & 14 & 0& 10 \\
    overcomplete QEC  & 28 & 0 & 14 & 0 &11\\
    logical T-gate            & 95 & 29 & 3 & 7 & 22 \\
    magic encoding    & 41 & 22 & 1 &0 &15 \\
    logical CS with subsequent reset       & 47 & 7 & 2 &7 &19 \\
    nnn reduced QEC           & 92 & 42 & 0 & 0 &17 \\
    nnn Encoding      & 29 & 1 & 0 & 0 &13 \\
   \hline
\end{tabular}
\caption{Number of two- and multi-qubit gates required for the different circuits with all-to-all connectivity ( row 1 to 5) and next-nearest neighbor (nnn) connectivity (row 6 and 7).}
\label{tab:resources}
\end{table}

\newpage
\bibliographystyle{apsrev4-2_limit_authors}

\bibliography{references}

\newpage

\end{document}